\def\PRL#1#2#3{{\sl Phys. Rev. Lett.} {\bf #1}, #2 (#3)}

\def\PLB#1#2#3{{\sl Phys. Lett.} {\bf B#1}, #2 (#3)}

\def\be{\begin{equation}}
\def\ee{\end{equation}}
\def\bea{\begin{eqnarray}}
\def\eea{\end{eqnarray}}
\def\bq{\begin{quote}}
\def\eq{\end{quote}}

\def\gappeq{\mathrel{\rlap {\raise.5ex\hbox{$>$}}
{\lower.5ex\hbox{$\sim$}}}}

\def\lappeq{\mathrel{\rlap{\raise.5ex\hbox{$<$}}
{\lower.5ex\hbox{$\sim$}}}}

\def\Rb{$R  \! \! \! \! /$}

\documentstyle [12pt,epsf,epsfig]{article}
\begin{document}
\thispagestyle{empty}
\begin{flushright}
{CERN-TH/96-258} \\
{CfPA-96-th-20} \\
{MPI-PhT-96-68} \\
{September 1996} \\
\end{flushright}
\vspace{1cm}
\begin{center}
{\large {\bf Basis-Independent Measures of $R$-Parity Violation }} \\
\vspace{.1cm}
\end{center}
\vspace{1cm}
\begin{center}
{\bf Sacha Davidson} \\
\vspace{.3cm}
{Max-Planck-Institut fur Physik\\
F\"{o}hringer Ring 6, 80805, M\"{u}nchen, Germany}\\
\vspace*{5mm}
and\\
\vspace*{5mm}
{\bf John Ellis}\\
\vspace{.3cm}
{Theory Division, CERN, CH 1211, Gen\`eve 23, Switzerland}
\end{center}
\hspace{3in}

\begin{abstract}

We construct basis-independent expressions
that measure the magnitude of $R$-parity breaking
due to possible superpotential terms
in the Minimal Supersymmetric extension of the Standard Model,
in the absence of soft supersymmetry-breaking terms and
spontaneous gauge symmetry breaking. We also
discuss briefly their application to a consistent treatment of
cosmological constraints on $R$-parity violation.
\end{abstract}
\vspace*{5cm}  
\begin{flushleft}
CERN-TH/96-258\\
September 1996
\end{flushleft}
\vfill\eject

\setcounter{page}{1}
\pagestyle{plain}

\section{Introduction}

Supersymmetry (SUSY) \cite{SUSY} is a popular extension of the
Standard Model, which renders natural its large mass
hierarchy: $M_P / M_W >> 1$.
Since the number of particles in even the Minimal
Supersymmetric extension of the Standard Model (MSSM)
is more than doubled, a number
of new tree-level interactions are allowed,  some of which
violate baryon and/or lepton number. 
A discrete symmetry called $R$ parity  is often
imposed to forbid all such interactions.
Alternatively, one may consider a broader class of
theories containing such interactions,
and constrain their coupling
constants to be small enough that they do not conflict
with experimental observations \cite{RPexpt} and
cosmological bounds \cite{cosm}.

The $R$-non-conserving interactions that
violate lepton number have the important feature that 
they can be moved around the Lagrangian by field redefinitions
\cite{RP}, introducing a possible ambiguity that should be
taken into account when evaluating these constraints.
This is not in general a serious problem with accelerator
bounds, because there it is clear that one is working in
the mass-eigenstate basis. However,
in cosmology the correct choice of basis is less clear,
because the mass-eigenstate basis changes
with temperature, and an interaction-eigenstate basis would
be more natural. A desirable resolution of this ambiguity
is to define  expressions that are
independent of the basis chosen for the fields to
measure the amount of $R$ violation, 
which is the purpose of this letter\footnote{Our approach
is similar in philosophy to the analysis by Jarlskog and others \cite{Jarl}
of basis-independent measures of CP violation in the Standard
Model.}.

In the Minimal Supersymmetric extension of the Standard
Model (MSSM) with $R$ conservation, the superpotential
can be written in the form
\be
W = \mu_H \bar{H} H + h_u^{ij} \bar{H} Q_i U^c_j + 
h_d^{ij} H Q_i D^c_j + h_e^{ij} H L_i E^c_j \label{S1}
\ee
where $i,j= 1,...,3$ are quark and lepton generation indices.
In addition to the supersymmetric interactions parametrized
by (\ref{S1}), one must introduce soft SUSY-breaking terms 
which include
\be
 {\rm mass ~ terms} + B_H \bar{H} H +
A_u^{ij} \bar{H} Q_i U^c_j + 
A_d^{ij} H Q_i D^c_j + A_e^{ij} H L_i E^c_j 
\label{soft}
\ee
as well as  gaugino mass terms. We use
capitalized fields to represent both the
superfields (as in eqn (\ref{S1})) and their scalar
components (as in eqn (\ref{soft})).
Equation (\ref{soft}) and the Lagrangian derived from
(\ref{S1}) are invariant under the multiplicatively-conserved
$Z_2$ global symmetry $R = 3B + L + 2S$,
where $B$ is the baryon number
of any field, $L$ is its lepton number, and $S$ is its spin.

Once we allow for interactions that do not conserve $R$,
we can add four more $B$- or $L$-violating terms to the superpotential:
\be
W_{R  \! \! \! \! / } = \epsilon_i H  L_i + \lambda_1^{ijk} L_iL_j E^c_k + 
\lambda_2^{ijk} L_iQ_jD^c_k + \lambda_3^{ijk} U^c_iD^c_jD^c_k \label{S2}
\ee
In this letter
we do not consider $\lambda_3$, and concentrate
on the $L$-violating interactions $\epsilon, \lambda_{1,2}$.
To appreciate the meaning of these terms, it is convenient to 
recall that the Higgs field $H$ and the lepton doublets $L_i$ have the same
gauge quantum numbers. In the Standard Model,
they are distinguished by the fact that
the Higgs is a scalar, and the leptons are fermions 
carrying global quantum number that are automatically conserved
if the neutrinos are massless. In the MSSM with the minimal
superpotential (\ref{S1}),
lepton number is still conserved. However, SUSY removes the
distinction between $H$ and the $L_i$ based on their spins, and
once we allow for $R$
violation, there is no unambiguous distinction between the
$H$ and the $L_i$, and we can assemble them in a vector
\be
\phi_I = (H, L_i) \label{phi}
\ee
whose index $I$ runs from 0 to 3. A Lagrangian contains
$R$ violation if different interactions choose
conflicting directions in $\phi$ space to be
the Higgs.

Combining (\ref{S1}) and (\ref{S2}), writing 
$\mu_I = (\mu ,\epsilon_i)$, 
$\frac{1}{2}h_e^{jk} = \lambda_e^{0jk}$, $\lambda_1^{ijk} = \lambda_e^{ijk},
\lambda_2^{ijk} = \lambda_d^{ijk}$ and
neglecting $\lambda_3$ as already mentioned, we may write
the superpotential as
\be
W= \mu^I \bar{H}  \phi_I + \lambda_e^{IJk} \phi_I \phi_J E^c_k + 
\lambda_d^{Ijk} \phi_IQ_jD^c_k  + h_u^{ij} \bar{H} Q_i U^c_j\label{S}
\ee
This parametrization could be extended to discuss 
soft SUSY-breaking terms, but we shall not do so in this paper.
The appearance of $R$ violation can  be moved  around
the Lagrangian by unitary transformations on the fields,
depending on what interaction one uses to define the
Higgs. One must therefore be careful to specify the basis 
in which one is 
working, or else discuss $R$ violation in a basis-independent 
formalism. 

In this letter, we opt for the latter,  and  construct 
combinations of coupling constants  which are
zero when $R$ is conserved, and which are independent
of the basis transformations in $\phi_I$ space. We normalise these
invariants relative to the magnitudes of the superpotential couplings
in (\ref{S}), so that they provide some measure of
the relative magnitude of $R$ violation present. We present
geometrical interpretations of these invariants, as well as
generic diagrams that give rise to them. We then
discuss briefly cosmological constraints on
$R$ violation, as an example how the invariants
can be used.  In this letter, we work in the absence of
soft supersymmetry breaking and
gauge symmetry breaking; a more complete analysis will be presented
elsewhere \cite{DE2}.

\section{Invariants in Simplified Models}

As warm-up exercises which help provide some intuition,
we first discuss two toy examples: first a one-generation
model, and then a model
with two generations of leptons, but only one 
generation of quarks.

Consider  a one-generation 
version of the superpotential (\ref{S}): this
means that upper-case indices run from 0 to 1, and
there are no lower-case indices. Since $\lambda_e^{IJ}$ is
antisymmetric in the indices $I,J$, there is no
$LLE^c$ interaction in this model, and the
Yukawa coupling $HLE^c$ is invariant under
rotations in $\phi$ space. However, by its presence,
the $\lambda_e^{IJ}$ interaction forces
one direction in $\phi$ space to be a lepton. 
The Higgs can be defined either from the $\mu^I \phi_I \bar{H}$
term, or from the $\lambda_d^I \phi_I Q D^c$ term. If, for instance,
the basis in $\phi$ space is chosen such that $\mu^1 = 0 $, then
the superpotential becomes
\be
W = \mu_H \bar{H} H + h_u \bar{H} Q U^c + 
h_d H Q D^c + h_e H L E^c  + \lambda_2 LQD^c 
\ee
where  the $R$ violation is contained in 
\be
\lambda_2 =  \frac{ \mu^{0*} \lambda_d^1 - \mu^{1 *}  \lambda_d^0}
{[ |\mu^0|^2 + |\mu^1|^2]^{1/2}}
\ee
Alternatively, one could choose a (primed) basis such
that $\left. \lambda_d^1 \right.' = 0$, but 
$\left.\mu^1 \right. ' = - ( \mu^{0*} \lambda_d^1 - \mu^{1 *}  \lambda_d^0 )/ 
[ |\lambda_d^0|^2 + |\lambda_d^1|^2]^{1/2}$.

\begin{figure}
\hglue 2.8cm
\epsfig{figure=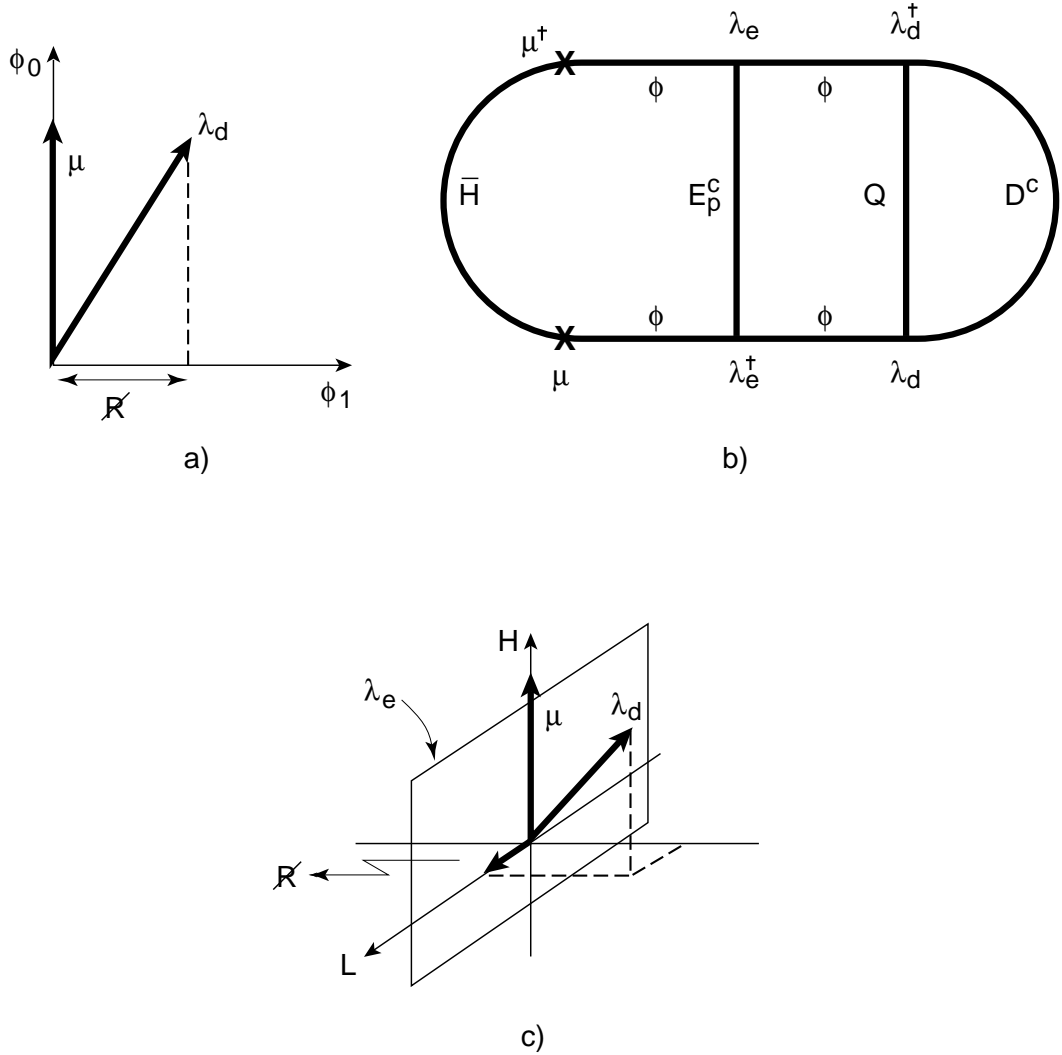,width=14cm}
\caption[]{A geometric interpretation of the invariant
\protect\( \mu^{\dagger} \lambda_e \lambda_d^{*} \protect\),
and the associated supergraph:  (a) is for
 a one generation model - $R$ parity is not conserved if \protect\( \mu
\protect\) and
\protect\( \lambda_d \protect\) do not choose the
same direction in \protect\( \phi \protect\) space to be the Higgs.
(b) The  supergraph corresponding to the invariant
\protect\( \delta_1 \protect\). With
multiple generations, there is a sum of diagrams, corresponding to
the sum over the right-handed lepton index \protect\( p \protect\).
 (c) A  geometric interpretation of 
\protect\(\mu^{\dagger} \lambda_e^p \lambda_d^{*}  \protect\)
in a two-generation model. The matrix \protect\( \lambda_e^p \protect\)
chooses a plane in \protect\( \phi \protect\) space, spanned
by a Higgs and a lepton. The projections of \protect\( \mu \protect\)
and \protect\( \lambda_d \protect\)
onto that plane each choose directions that should be the Higgs -
there is $R$ violation if these directions are not parallel.} 
\label{f1}
\end{figure}

It is clear that $R$ parity is broken in
this model if two conditions are satisfied.
Firstly, $\lambda_e$  must be present, so that
one direction in $\phi$ space is a lepton,
and secondly $\mu$ and $\lambda_d$ must choose different
directions in $\phi$ space to be the Higgs. A measure
of $R_p$ violation is then the length of the component
of $\lambda_d$ that is orthogonal to $\mu$: 
$ \mu^{0*} \lambda_d^1 - \mu^{1 *}  \lambda_d^0$, as
illustrated in Fig.~\ref{f1}(a).
It is helpful to normalize this measure
by dividing it by the lengths of the vectors $\mu$ and 
$\lambda_d$: 
\be
\epsilon = \frac{ \mu^{0*} \lambda_d^1 - \mu^{1 *}  \lambda_d^0}
                 { |\mu| |\lambda_d|}
\ee
The basis-dependent $R$-violating
coupling $\left. \mu^1\right.'$ is then $|\mu| \times \epsilon$,
and similiarly $\lambda_2 = |\lambda_d| \times \epsilon$.

Another way to express  the invariant measure of $R$
violation in this model is 
\be
\epsilon = \frac{ \mu^{I*} \lambda_e^{IJ} \lambda_d^{J*} }
 { |\mu| |\lambda_e| |\lambda_d|}
\ee
where $|\lambda_e|^2 = \lambda_e^{IJ} \lambda_e^{IJ*}$. This is
the same expression as previously discussed, but has the
advantage of including also the Yukawa coupling, which must be present
for the model to violate $R$. We prefer this formulation,
because it can readily be generalized to more generations. 

In fact, it is more appropriate to use the square of this expression as 
a measure of $R$ violation, because it corresponds
to a closed-loop Feynman diagram (see Fig.~\ref{f1}(b)),
such as would appear in a zero temperature
rate calculation.
We  therefore advocate
\be
\delta_1 = \frac{| \mu^{I*} \lambda_e^{IJ} \lambda_d^{J*}|^2 }
 { |\mu|^2 |\lambda_e|^2 |\lambda_d|^2}
\ee
as the invariant measure of $R$ violation in this 
simple one-generation model.

We now discuss the second toy model. It has two
lepton generations, so the matrices $\lambda_e^{IJp}$ 
enter in a non-trivial way, but the $\phi$ space is three-dimensional,
so that it is easy to understand geometrically
the invariants found below.
Suppose one has a superpotential of the form
(\ref{S}), with upper case indices that
run from 0 to 2 and lower case lepton indices that run from 1 to 2. 
The coupling constants define a number of directions
in $\phi$ space: $\mu$ and $\lambda_d$ correspond to
directions that one could choose to be the Higgs,
and  each $\lambda_e$  determines a plane spanned by a
Higgs and a lepton. The intersection of the two planes chooses
a direction for the Higgs. Each plane can therefore
be written as the cross product
of the Higgs and an orthogonal vector corresponding to the lepton:
\be
\lambda^{IJp} = L^{Ip} H^J - H^I L^{Jp}
\ee
As vectors in $\phi$ space, the $L^{Ip}$ depend
on the choice of basis for the right-handed leptons.
To see this, imagine choosing the Higgs 
to correspond to the direction $\phi^0$; the index
$I$ on $L^{Ip}$ then runs from 1 to 2, like $p$, and 
$L^{Ip}$ has the same form as the $2 \times 2$ Yukawa matrix.
It can be diagonalised by independent unitary
transformations on the left- and right-handed fields
(i.e., on the $I$ and $p$ indices), which means that the two
$L^{Ip}$, regarded as vectors in $\phi^I$ space,  are
orthogonal only for a specific choice of basis for the
$p$ indices.

This  model violates $R$ parity if the direction
chosen by one interaction to be the Higgs 
has a non-zero inner product 
with the direction chosen by
another interaction to be a lepton.
There are therefore three
scalars one can construct that measure \Rb \ .
If one projects $\mu$ onto the plane associated
with one of the Yukawa couplings, it picks out the
direction in the plane to be identified
with the Higgs. The orthogonal direction
is therefore a lepton, and its inner product with 
$\lambda_d$: $\mu^I \lambda_e^{IJ p} \lambda_d^J$,
is a measure of $R$ violation, see  Fig.~\ref{f1}(c).
This expression is not yet completely satisfactory, because
it carries a right-handed lepton index $p$, and so is not fully
basis-independent. This defect can be remedied
by multiplying it by its complex conjugate, and
summing over $p$:
\be
\delta_1 = \frac{ (\mu^{I*} \lambda_e^{IJ p} \lambda_d^{J*})
(\mu^{M} \lambda_e^{MN p*} \lambda_d^{N})}{|\mu|^2 |\lambda_e|^2 |\lambda_d|^2}
\ee
This measure of $R$ violation is invariant under all basis transformations
and corresponds to the closed supergraph 
shown in Fig.~\ref{f1}(b). However, we should emphasize 
that one disadvantage of summing over 
right-handed lepton indices is that we now only have a measure
of total $L$ violation, and are not able to determine whether
one of the individual $L_i$ might be conserved. 

\begin{figure}
\hglue 3.5cm
\epsfig{figure=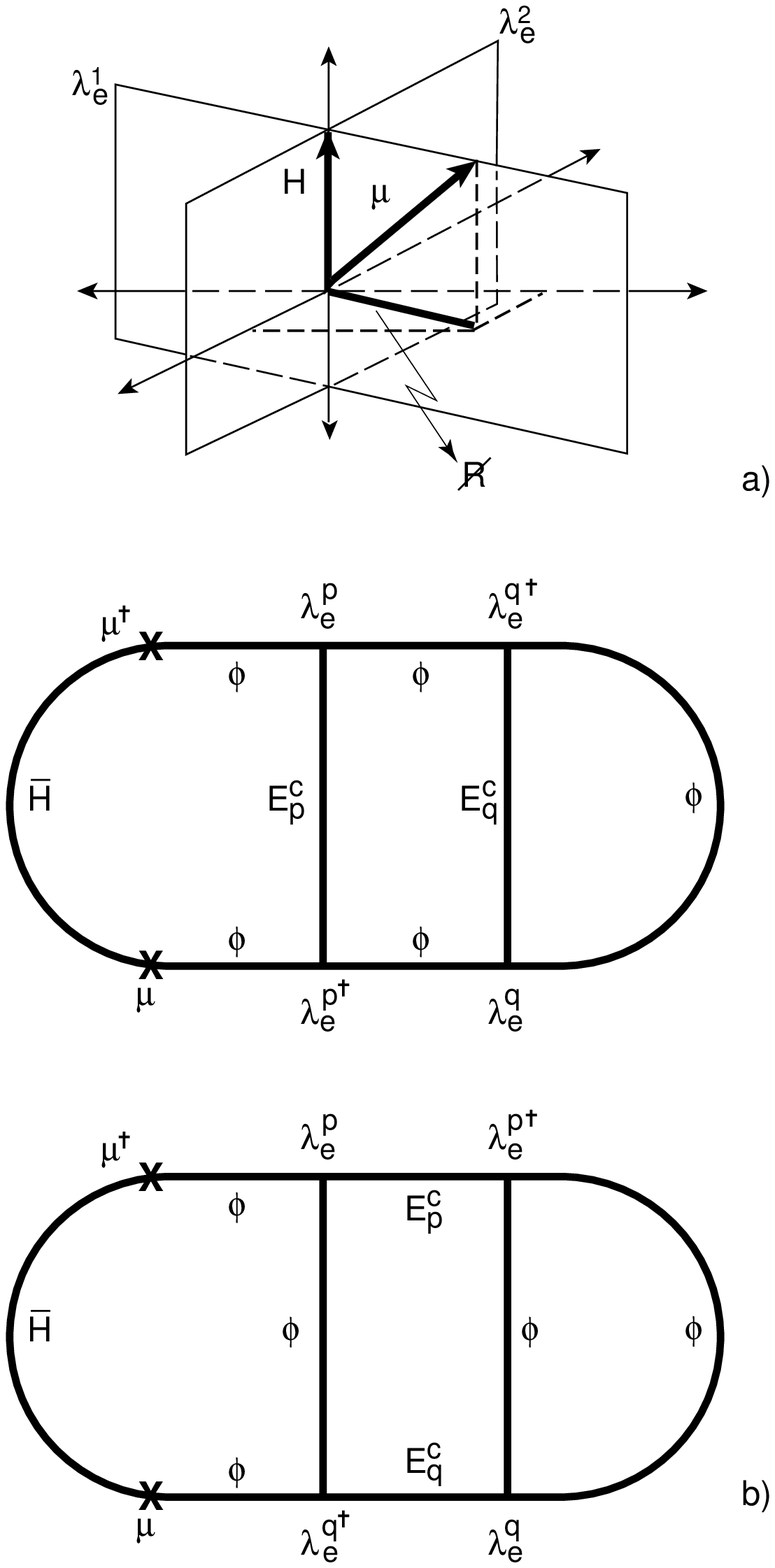,width=10cm}
\caption[]{(a) 
A  geometric interpretation of the invariant \protect\( \delta_2
\protect\) in a two-generation model, 
before the sum over right-handed lepton indices. Each
matrix \protect\( \lambda_e^p \protect\) corresponds to a plane 
in \protect\( \phi \protect\) space. The intersection of the planes
corresponds to the direction that is chosen to be the Higgs by the
\protect\( \lambda_e \protect\) matrices. If this direction  is not parallel
to the direction chosen by \protect\( \mu \protect\), 
there is R-parity violation. (b) The
supergraphs for  \protect\( \delta_2 \protect\):  
the right-handed lepton indices \protect\( p \protect\) 
and \protect\( q \protect\) are summed.}
\label{f2}
\end{figure}

\begin{figure}
\hglue 3.5cm
\epsfig{figure=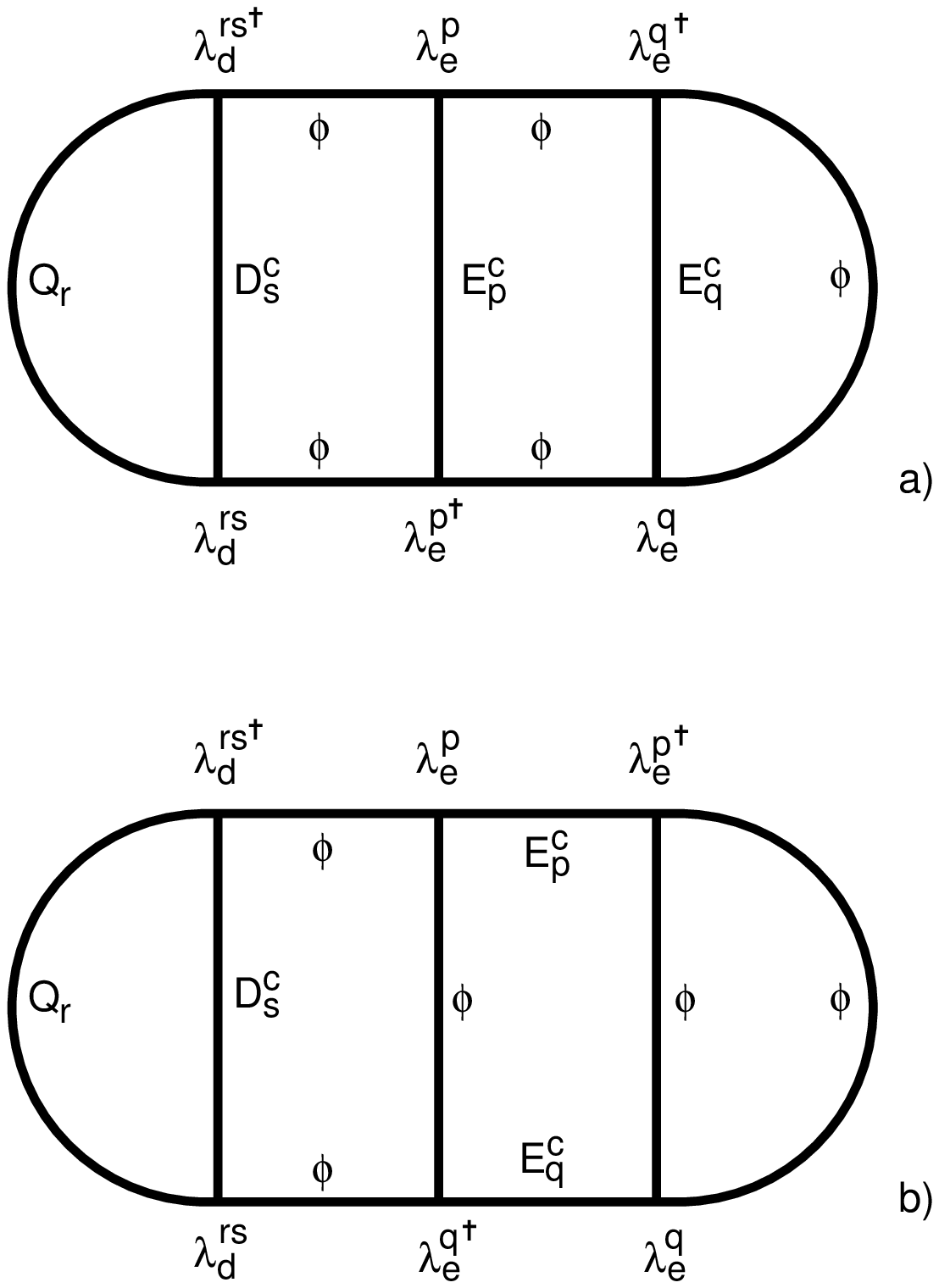,width=10cm}
\caption[]{ Supergraphs corresponding to the 
invariant \protect\( \delta_3 \protect\):   
the right-handed lepton indices \protect\( p \protect\) 
and \protect\( q \protect\) and the quark indices \protect\( r \protect\) 
and \protect\( s \protect\) are summed.}
\label{f3}
\end{figure}

We see also that $R$ is not conserved if the direction  
chosen to be the Higgs  by the two Yukawa couplings has
a component perpendicular to   $\mu$ or $\lambda_d$,
as seen in Fig.~\ref{f2}(a).
Let us consider $\mu$ (the invariant for $\lambda_d$ will
be a straightforward modification). We wish to know
the length of the component of $\mu$ in the plane spanned by
the two leptons $L^{p}$. It is clear that, up to an overall
normalization, this is
\be
|L^1|^2 |\mu \cdot L^2 |^2 + | L^2|^2 |\mu \cdot L^1|^2 - 2 (\mu \cdot L^1)
(L^1 \cdot L^2) (\mu \cdot L^2)
\ee
where the $\phi$ indices are suppressed. Note that $L^1$ and
$L^2$ are not orthogonal in a generic basis.  Writing this
expression in terms of the matrices $\lambda_e^{IJp}$, we get 
\be
\delta_2 = \frac{ \mu^{I *} \lambda_e^{IJp} \lambda_e^{JKq *} 
  \lambda_e^{KLq} \lambda_e^{LMp *} \mu^M
  -1/2 ( \mu^{I *} \lambda_e^{IJp} \lambda_e^{JKq *} \mu^K  \lambda_e^{LMp *}
  \lambda_e^{LMq} )}{|\mu|^2 Tr[ \lambda_e^{p*} \lambda_e^p \lambda_e^{q*} 
\lambda_e^q]}  
\label{compl1}
\ee
which corresponds to the difference between
the supergraphs of Fig.~\ref{f2}(b). The trace is over the
capitalised indices, and the lower case indices
are also summed. A similar expression for
$\lambda_d$, corresponding to the difference between the supergraphs
of Fig.~\ref{f3}, gives  
\be
\delta_3 = \frac{ \lambda_d^{Irs *} \lambda_e^{IJp} \lambda_e^{JKq *}
\lambda_e^{KLq} 
  \lambda_e^{LMp *} \lambda_d^{Mrs}
  -1/2 ( \lambda_d^{Irs *} \lambda_e^{IJp} \lambda_e^{JKq *} \lambda_d^{Krs}
  \lambda_e^{LMp *} \lambda_e^{LMq} )}{|\lambda_d|^2 
  Tr[ \lambda_e^{p*} \lambda_e^p \lambda_e^{q*} 
  \lambda_e^q]}  
  \label{compl2}
\ee
which completes our enumeration of measures of $R$ violation
in this second toy model, to this order in the couplings. 

\section{Three-Generation Model}

We are now in a position to generalize to 
the physical situation of three lepton
generations. In this case, not only are
the Yukawa coupling matrices $\lambda_e^{IJp}$ non-invariant
under $\phi$ basis transformations, but they can also contain
$R$ violation independently of the other interactions
present. In the limit where there
is no \Rb , they define a plane in $\phi$ space
that is spanned by the Higgs and a lepton.
The $\mu^I$ and $\lambda_d^{Ijp}$ again
define directions in $\phi$ space that one would
like to interpret as Higgs fields.  In the absence of
$R$ violation, these two directions
coincide with the Higgs direction chosen
by the planes of the three leptonic
Yukawa matrices $\lambda_e^{IJp}$. 

The simultaneous presence of at least
three interactions is required 
in order to provide $R$ violation. 1) At least
one Yukawa coupling must be present, so as to force one plane
in $\phi$ space to contain a lepton and a Higgs.
2) The second interaction can choose which direction
in this plane is the Higgs. 3) The third
interaction must then specify a direction
that conflicts with this choice.
Since one must square the amplitude in order to obtain a rate,
our invariants involve at least six coupling constants.

For simplicity, we do not consider expressions involving more than
the minimum three coupling constants required to obtain $R$ violation.
The list we now present consists of all the invariants 
that can be constructed out of three coupling constants,
assuming that at least one of them is a $\lambda_e$ and
summing over all three generations.
This list is complete, in the sense
that at least one of the invariants  is non-zero if
the theory contains $R$ violation at this order.

\begin{figure}
\hglue 3.5cm
\epsfig{figure=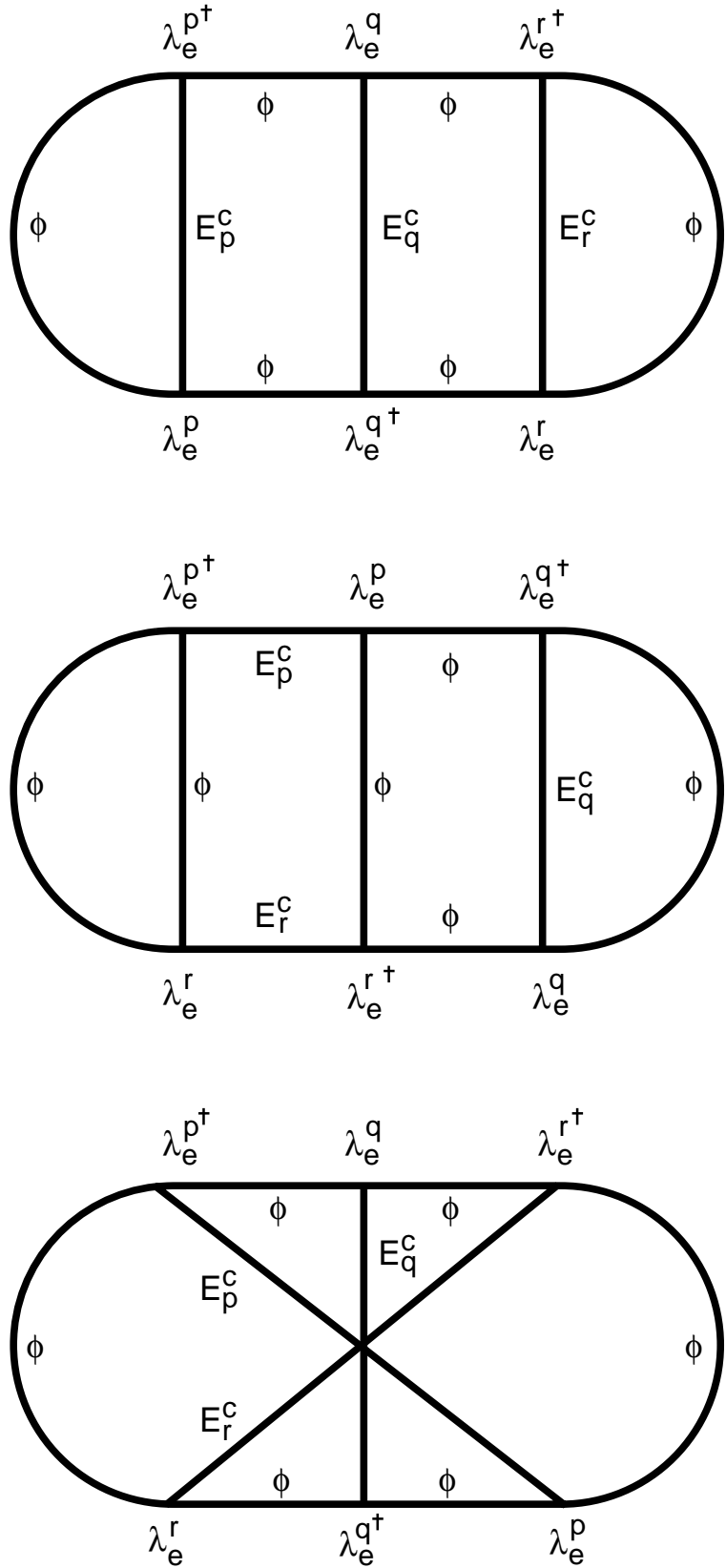,width=10cm}
\caption[]{ Supergraphs corresponding to the 
invariant \protect\( \delta_4 \protect\):  
the right-handed lepton indices \protect\( p \protect\), 
\protect\( q \protect\) and \protect\( r \protect\) are summed.}
\label{f4}
\end{figure}

\begin{figure}
\hglue 3.5cm
\epsfig{figure=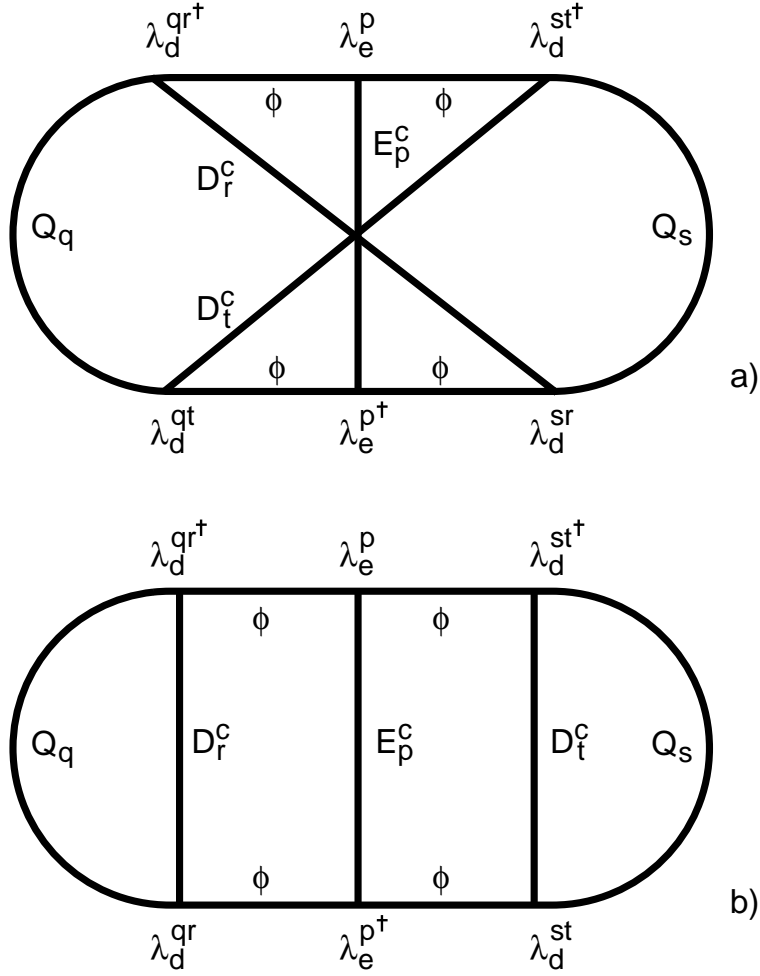,width=10cm}
\caption[]{ Supergraphs corresponding to the 
invariant \protect\( \delta_5 \protect\):  
the right-handed lepton index \protect\( p \protect\) 
and the quark indices \protect\( q \protect\), 
\protect\( r \protect\),  \protect\( s \protect\) 
and \protect\( t \protect\) are summed.}

\label{f5}
\end{figure}

In an exactly supersymmetric theory,
the following squares of combinations of three 
coupling-constant matrices  are invariant under 
transformations in $\phi$ space, and zero in
an $R$-conserving theory\footnote{They also give the above
two-generation expressions in the limit where
one generation's interactions are set to zero.}:
\be
\delta_1 = \frac{(\mu^{I*} \lambda_e^{IJp} \lambda_d^{Jqr*})
(\mu^{K} \lambda_e^{KLp*} \lambda_d^{Lqr})}
 {|\mu|^2 |\lambda_e|^2 |\lambda_d|^2} \label{sug1}
\ee
\be
\delta_2 = \frac{\mu^{I *} \lambda_e^{IJp} \lambda_e^{JKq *} 
  \lambda_e^{KLq} \lambda_e^{LMp *} \mu^M
  -1/2 ( \mu^{I *} \lambda_e^{IJp} \lambda_e^{JKq *} \mu^K  \lambda_e^{LMp *}
  \lambda_e^{LMq} )}
{|\mu|^2 Tr[ \lambda_e^{p*} \lambda_e^p \lambda_e^{q*} 
\lambda_e^q]}  
 \label{sug2}
\ee
\be
\delta_3= \frac{\lambda_d^{Irs *} \lambda_e^{IJp} \lambda_e^{JKq *} 
  \lambda_e^{KLq} \lambda_e^{LMp *} \lambda_d^{Mrs}
  -1/2 ( \lambda_d^{Irs *} \lambda_e^{IJp} \lambda_e^{JKq *} \lambda_d^{Krs}
  \lambda_e^{LMp *} \lambda_e^{LMq} ) }
{|\lambda_d|^2 Tr[ \lambda_e^{p*} \lambda_e^p \lambda_e^{q*} 
\lambda_e^q]}  
 \label{sug3}
\ee

\be
\delta_4 = \frac{Tr 
[\lambda_e^p \lambda_e^{q*} \lambda_e^r \lambda_e^{r*} \lambda_e^{q} 
\lambda_e^{p*}]
+ Tr 
[\lambda_e^p \lambda_e^{q*} \lambda_e^r \lambda_e^{p*} \lambda_e^{q} 
\lambda_e^{r*}]
 - Tr 
[\lambda_e^p \lambda_e^{q*} \lambda_e^r \lambda_e^{p*}] Tr[ \lambda_e^{q} 
\lambda_e^{r*}]}
{ Tr 
[\lambda_e^p \lambda_e^{p*} \lambda_e^r \lambda_e^{r*} \lambda_e^{q} 
\lambda_e^{q*}]
} \label{sug4}
\ee 

\be
\delta_5 = \frac{(\lambda_d^{Ist*} \lambda_e^{IJp} \lambda_d^{Jqr*})
(\lambda^{Kst} \lambda_e^{KLp*} \lambda_d^{Lqr})
+ (\lambda_d^{Ist*} \lambda_e^{IJp} \lambda_d^{Jqr*})
(\lambda_d^{Kqt} \lambda_e^{KLp*} \lambda_d^{Lsr})}
 { |\lambda_e|^2 |\lambda_d|^4} \label{sug5}
\ee
The traces are over the capitalized $\phi$ indices.
The lower-case indices correspond to quark and
right-handed lepton generations, and are also summed.
The numerators correspond to the closed supergraphs
of Figures \ref{f1}, \ref{f2}, \ref{f3}, \ref{f4} and \ref{f5},
respectively. The magnitudes of the matrices and vectors 
in the denominators of the above expressions are defined by
\be
|\mu|^2  = \sum_I \mu_I \mu_I^* \label{mag1}
\ee
\be
|\lambda_d|^2 = \sum_{Ipq} \lambda_d^{Ipq} \lambda_d^{Ipq *} \label{mag2}
\ee
\be  
|\lambda_e|^2 =  \sum_{IJr} \lambda_e^{IJr} \lambda_e^{IJr *}  \label{mag3}
\ee
There are two new invariants in this three-generation model, shown in
Figures~\ref{f4} and \ref{f5}.
The Yukawa couplings $\lambda_e$ now contain $R$ violation all by
themselves, which is measured by $\lambda_4$. It is also possible that
the various $\lambda_d$ (there are nine of them in the case of three
quark generations) might choose different
directions in $\phi$ space to be the Higgs: $\delta_5$
measures the component of $\lambda_d^{st}$ perpendicular to
$\lambda_d^{qr}$. 

One can see by inspection that these combinations 
are zero in an $R$-conserving theory, and contain the minimal 
(because the invariants correspond to squares of matrix elements)
two powers of the $R$-violating coupling constants 
from equation (\ref{S2}).

\section{Cosmological Applications of the Invariants}

The uses of the invariants we have defined above are
limited by the fact that we have neglected soft supersymmetry breaking
and have not included the Higgs vev. However, these
restrictions do not hinder application in the early Universe
above the electroweak phase transition (EPT). 
A limitation that is more serious for this
application is that we have also
summed over the quark and right-handed lepton indices.
We plan to address this and the previous
deficiencies in a subsequent
publication \cite{DE2}. 
However, we can already use the above invariants 
to discuss cosmological constraints on $R$ violation,
providing we restrict ourselves to sums over quarks and
right-handed leptons. We will first review
these cosmological constraints, then calculate 
them in a one-generation model, where there are
no unwanted sums, and finally discuss the constraints
on our invariants in the three-generation case.

In many cosmological scenarios, the survival  of the baryon asymmetry
can provide strong bounds on $R$ violation. If the asymmetry
was created before the EPT, it would have
to contend with the anomalous electroweak $B+L$-violating processes
known to be present in the thermal bath before the EPT. 
For a baryon asymmetry to survive, there would have to be a $B-L$
asymmetry.
If $R$-violating interactions were also present in the thermal
bath, the asymmetry would be washed out if they took $B-L$ to zero.
Requiring all $R$-violating
interaction rates to be less than the expansion rate, so that
they are unable to perpetrate this catastrophe,
may yield strong bounds on the coupling constants
of (\ref{S2}). However, there are two
loopholes in this argument: the baryon asymmetry may be
made at the EPT, or the $R$-violating
interactions that are in thermal equilibrium may
only wash out one lepton number, e.g., $ L_1$, and not
wash out $B-L$. For the purposes of this section, we neglect these
loopholes.

We require all $R$-violating interactions
to be out of thermal equilibrium at $T \sim 100$ GeV,
which we assume to be of the same order as the supersymmetric
particle masses and the EPT temperature. 
We assume that the supersymmetric particles are present with a density
$\sim T^3$ in the thermal bath, and neglect mass effects, whose
inclusion would change the bounds only slightly~\cite{cosm}. 
The supersymmetry-breaking
scalar masses are not present in the invariants, so it is
consistent to neglect the contribution of
these masses to $R$-violating rates. 

As an example of the estimation of such a rate, we consider
$\delta_1$ (\ref{sug1}), first in the
one-generation model, and then with three generations. In the
one-generation case, it is clear that if all of
the rates associated with the interactions $\mu$, $\lambda_e$ and
$\lambda_d$ are not in equilibrium, i.e.,
greater than the expansion rate $H$, 
the corresponding coupling constant 
can be effectively neglected. 
In fact, all the interactions
of the MSSM are in thermal equilibrium before
the EPT, so this condition is satisfied.  In 
the one-generation toy model, the $R$-violating rate is 
\be
\Gamma_{R \! \! \! \! /} \simeq \delta_1 \times \Gamma_{min} 
\label{del1rate}
\ee
where $\Gamma_{min}$ is the smaller of the rates
associated with $|\mu|$ and $|\lambda_d|$.
As discussed in section 2, the $R$-violating
coupling constants in  the 
one-generation model are $\mu^1$ and/or $\lambda_d^1$:
$\lambda_e$ enters into $\delta_1$ only to
ensure that some direction  in $\phi$ space is a lepton.
We constrain (\ref{del1rate}) to be out of equilibrium. 

The rates associated with $|\mu|$ and $|\lambda_d|$ 
are estimated~\cite{cosm} to be:
\be
\Gamma_{\mu} \simeq 10^{-2} \frac{ \mu^2}{T}, ~~~~~~~\Gamma_{\lambda_d}
\simeq 10^{-2} \lambda_d^2 T
\ee
Taking $\mu \sim 100$ GeV, $\lambda_d \sim h_b \sim 3 \times 10^{-2}$,
the bound $\delta_1 \Gamma_{\lambda_d} < H$ yields
\be
\frac{ ( \mu^1 \lambda_d^0 - \lambda_d^1 \mu^0)^2}{|\mu |^2} < 10^{-11}  
\label{star}
\ee
If we assume self-consistently that $R$ violation 
is small, we can
work in the MSSM thermal 
mass eigenstate basis, as is appropriate just
before the electroweak phase transition.  In
this basis  (\ref{star}) gives the bounds
\be
\frac{ \mu^1}{\mu^0} < 3  \times 10^{-6} , ~~~~~\lambda_d^1 < 10^{-7}
\label{27}
\ee
Note that this bound on $\mu^1$ is 
considerably weaker than what one would obtain naively
by requiring  the  rate associated with a mass term \cite{cosm}
to be out of equilibrium:
\be
\Gamma \sim 10^{-2} \left( \frac{\mu^i}{\mu^0} \right)^2 T \lappeq H
\label{old}
\ee 
which would give $\mu^i \lappeq 10^{-5}$ GeV.

In our invariant approach, it is easier to understand when
$R$ violation is (or is not) in equilibrium in the
early Universe. In the basis-dependent approach, one can
move the $R$ violation 
from the mass terms to the Yukawa terms. Interaction
rates associated with masses and Yukawa couplings scale differently
with temperature,  so
the temperature at which the $R$ violation
goes in or out of equilibrium appears to depend on
the basis one chooses. This confusion is avoided
by noting that all the interactions that enter into
the definition of  an invariant must be
in equilibrium, as well as the invariant 
times one of the rates, e.g., (\ref{del1rate}). 

In the case of
three generations, the invariants we have defined  cannot
be used to estimate rates so easily, because we have
summed over the quark and right-handed lepton-generation
indices.  These sums made the invariants
independent of basis transformations on all the
fields. However, since it is only the
rotations in $\phi$ space that move the $R$ violation
around, for practical purposes
we can drop the sums over the quark and right-handed
lepton indices. The ``invariants'' are
then basis-dependent, and we compute them in
the thermal mass-eigenstate basis for  quarks
and right-handed leptons, which should be the right basis
in which to compute rates in the early Universe.
In principle, it would be better to quote bounds on the ``invariants'',
as in (\ref{star}),rather than on terms from the sum making them up, as in
(\ref{27}) because
there could be cancellations between the different
contributions. However, assuming such cancellations do not
take place, the bounds we now present are 
easier to interpret.  

We consider again, as an example, $\delta_1$. In the
absence of the sums over lower-case indices, 
we now have 27 invariants: there
are three $\lambda_e$, and  nine $\lambda_d$, because the leptonic
Yukawa-coupling matrices are diagonal in the thermal mass-eigenstate
basis, but the quark Yukawa couplings are not. The $R$-violating
entries in $\lambda_e$ do not contribute to $\delta_1$
at lowest order, so we neglect them.  In the mass-eigenstate
basis, $\lambda_e^p$ will therefore pick out the plane in
$\phi$ space spanned by the Higgs and the lepton $L^p$. 
This case now resembles the one-generation toy model
that we considered previously, except that now
there are nine possible vectors  $\lambda_d^{rs}$ in $\phi$ space.
For each $r$ and $s$, we have 
\be
\frac{( \mu^p \lambda_d^{0rs} - \mu^0 \lambda_d^{prs})^2}
{ |\mu |^2 | \lambda_d^{rs}|^2} 
\times 10^{-2} \, | \lambda_d^{rs}|^2 T < H
\ee
at $T \sim 100$ GeV. If one chooses to work in the
MSSM thermal mass-eigenstate basis, i.e., assume that $R$ violation
is small, one gets the following generic basis-dependent bounds:
\be
\frac{\mu^p}{\mu^0} < 3 \times 10^{-6} ~, ~~~~ \lambda_d^{prs} < 10^{-7}
\label{what}
\ee
where $p,r,s$ run from 1 to 3, and we have assumed that
$\lambda_d^{011} \simeq h_b \simeq 3 \times 10^{-2}$, where
$h_b$ is the bottom-quark Yukawa coupling.

Similarly, we do not sum over the quark and
right-handed lepton indices in the invariants 
$\delta_2, \delta_3, \delta_4 $ and $\delta_5$,
but instead compute ``invariants'' with the quarks and 
right-handed leptons in the thermal mass-eigenstate basis.
Requiring that the rates associated with
these ``invariants'' be smaller than the expansion rate of
the Universe at $ T \sim 100$ GeV, 
we obtain the generic bound
\be
\lambda \lappeq  10^{-7}
\ee
where $\lambda$ is some $R$-violating coupling in
the MSSM thermal mass-eigenstate basis. We also get a bound on
$\mu^i/\mu^0$ from
$\delta_2$, but it is weaker than (\ref{what}) because the
bottom-quark Yukawa coupling is larger than that of the
$\tau$. 

Our generic bounds on the $R$-violating dimensionless
coupling constants $\lambda_1, \lambda_2$ (see  equation (\ref{S2}))
are the same as previously calculated \cite{cosm}. The constraint on
$\mu^i$ is weaker, because the bound had previously been estimated by
requiring that $( \mu^1 / \mu^0 )^2 \alpha T < H$.
However, since the gauge interactions are diagonal in $\phi$ space, they do not
mix the different $\phi^I$, so the bound on $\mu^i$ comes from the mass
corrections to the Yukawa interactions, as demonstrated by our
analysis of invariants\footnote{Note that this would not be
the case for cosmological bounds on an 
$R$-violating soft supersymmetry-breaking mass. In this case, 
the mass-eigenstate bases for the fermions and scalars are different,
so the D-term interactions violate $R$ and the bound on the
$R$-violating soft supersymmetry-breaking masses would be of order
(\ref{old}). If $\mu^i$ is rotated away in a theory
with soft supersymmetry-breaking masses, it generates $R$-violating
scalar masses proportional to soft mass differences,
so the bound (\ref{old}) would still be overly optimistic, assuming
that these mass differences are small.}.

\section{Conclusions}

We have presented in this paper an analysis of $R$ violation in 
supersymmetric models that is formulated in terms of
basis-independent measures, in analogy with the treatment of
CP violation in~\cite{Jarl}. The enumeration of these measures
is quite complicated in general, and we have presented here a
simplified analysis in which soft supersymmetry-breaking masses
and Higgs v.e.v.'s are neglected. This restriction still permits
$R$-violating rates in the early Universe to be discussed, and
we have presented an analysis of the generic bounds obtained in
one- and three-generation models, using all the invariant
measures of $R$-violating rates that appear up to sixth order in the
couplings. Our upper limits on generic Yukawa couplings are
similar to those obtained previously~\cite{cosm}, but our bounds on the
supersymmetric mass-mixing parameters $\mu^i$ are less restrictive.

We have not studied in this paper loopholes in these cosmological
bounds, which may be exploited if the baryonic asymmetry is
generated at the EPT, or if there is a conserved quantum number or
flavour symmetry that preserves a pre-existing baryon asymmetry, or
if there are cancellations among the different contributions to our
invariants. These and the inclusion of soft supersymmetry breaking
and Higgs v.e.v.'s remain to be analyzed in future work~\cite{CDEO4}.

\subsection*{Acknowledgements} We thank Ralf Hempfling, Hitoshi Murayama
and Michael Peskin for useful discussions, and the Berkeley Center for
Particle Astrophysics and Lawrence Berkeley National Laboratory for
their kind hospitality during the course of this work.

\newpage

\end{document}